\documentclass[prl,aps,twocolumn,showpacs,bibnotes,superscriptaddress,epsf]{revtex4}%superscriptaddress,
\usepackage{graphicx}% Include figure files
\usepackage{dcolumn}% Align table columns on decimal point
\usepackage{bm}% bold math
\usepackage{SIunits}
\usepackage{tabularx}

\begin{document}

\title{Magnetoresistance evidence on surface state and field-dependent bulk gap in Kondo insulator SmB$_6$}
\author{F. Chen}\affiliation{Hefei National Laboratory for Physical Science at Microscale and Department of Physics, University of Science and Technology of China, Hefei, Anhui 230026, People's Republic of China}\affiliation{Key Laboratory of Strongly-coupled Quantum Matter Physics, University of Science and Technology of China, Chinese Academy of Sciences, Hefei 230026, China}
\author{C. Shang}\affiliation{Hefei National Laboratory for Physical Science at Microscale and Department of Physics, University of Science and Technology of China, Hefei, Anhui 230026, People's Republic of China}\affiliation{Key Laboratory of Strongly-coupled Quantum Matter Physics, University of Science and Technology of China, Chinese Academy of Sciences, Hefei 230026, China}
\author{Z. Jin}\affiliation{Wuhan National High Magnetic Field Center (WHMFC), Huazhong University of Science and Technology, Wuhan, Hubei 430074, China}
\author{D. Zhao}\affiliation{Hefei National Laboratory for Physical Science at Microscale and Department of Physics, University of Science and Technology of China, Hefei, Anhui 230026, People's Republic of China}\affiliation{Key Laboratory of Strongly-coupled Quantum Matter Physics, University of Science and Technology of China, Chinese Academy of Sciences, Hefei 230026, China}
\author{Y. P. Wu}\affiliation{Hefei National Laboratory for Physical Science at Microscale and Department of Physics, University of Science and Technology of China, Hefei, Anhui 230026, People's Republic of China}\affiliation{Key Laboratory of Strongly-coupled Quantum Matter Physics, University of Science and Technology of China, Chinese Academy of Sciences, Hefei 230026, China}
\author{Z. J. Xiang}\affiliation{Hefei National Laboratory for Physical Science at Microscale and Department of Physics, University of Science and Technology of China, Hefei, Anhui 230026, People's Republic of China}\affiliation{Key Laboratory of Strongly-coupled Quantum Matter Physics, University of Science and Technology of China, Chinese Academy of Sciences, Hefei 230026, China}
\author{Z. C. Xia}\affiliation{Wuhan National High Magnetic Field Center (WHMFC), Huazhong University of Science and Technology, Wuhan, Hubei 430074, China}
\author{A. F. Wang}\affiliation{Hefei National Laboratory for Physical Science at Microscale and Department of Physics, University of Science and Technology of China, Hefei, Anhui 230026, People's Republic of China}\affiliation{Key Laboratory of Strongly-coupled Quantum Matter Physics, University of Science and Technology of China, Chinese Academy of Sciences, Hefei 230026, China}
\author{X. G. Luo}\affiliation{Hefei National Laboratory for Physical Science at Microscale and Department of Physics, University of Science and Technology of China, Hefei, Anhui 230026, People's Republic of China}\affiliation{Key Laboratory of Strongly-coupled Quantum Matter Physics, University of Science and Technology of China, Chinese Academy of Sciences, Hefei 230026, China}\affiliation{Collaborative Innovation Center of Advanced Microstructures, Nanjing University, Nanjing, 210093, China}
\author{T. Wu}\email{wutao@ustc.edu.cn}\affiliation{Hefei National Laboratory for Physical Science at Microscale and Department of Physics, University of Science and Technology of China, Hefei, Anhui 230026, People's Republic of China}\affiliation{Key Laboratory of Strongly-coupled Quantum Matter Physics, University of Science and Technology of China, Chinese Academy of Sciences, Hefei 230026, China}\affiliation{Collaborative Innovation Center of Advanced Microstructures, Nanjing University, Nanjing, 210093, China}
\author{X. H. Chen}\email{chenxh@ustc.edu.cn}\affiliation{Hefei National Laboratory for Physical Science at Microscale and Department of Physics, University of Science and Technology of China, Hefei, Anhui 230026, People's Republic of China}\affiliation{Key Laboratory of Strongly-coupled Quantum Matter Physics, University of Science and Technology of China, Chinese Academy of Sciences, Hefei 230026, China}\affiliation{Collaborative Innovation Center of Advanced Microstructures, Nanjing University, Nanjing, 210093, China}

\begin{abstract}

Recently, the resistance saturation at low temperature in Kondo insulator SmB$_6$, a long-standing puzzle in condensed matter physics, was proposed to originate from topological surface state. Here, we systematically studied the magnetoresistance of SmB$_6$ at low temperature up to 55 Tesla. Both temperature- and angular-dependent magnetoresistances show a similar crossover behavior below 5~K. Furthermore, the angular-dependent magnetoresistance on different crystal face confirms a two-dimensional surface state as the origin of magnetoresistances crossover below 5K. Based on two-channels model consisting of both surface and bulk states, the field-dependence of bulk gap with critical magnetic field (H$_c$) of 196~T is extracted from our
temperature-dependent resistance under different magnetic fields. Our results give a consistent picture to understand the low-temperature transport behavior in SmB$_6$, consistent with topological Kondo insulator scenario.

\end{abstract}


\maketitle

Mixed-valent compound SmB$_6$ with CsCl-type structure is a well-known Kondo
Insulator (KI)\cite{Menth, Varma}. At high temperature, the 4f
electrons of Sm ions exhibit local character and are decoupled from
5d conduction bands. With temperature decreasing across the Kondo
temperature (T$_K$), the itinerant character of 4f electrons shows up
due to the hybridization between 4f electrons and 5d conduction
bands. Meanwhile, a small energy gap emerges at the Fermi
level (E$_F$) which will lead to the divergence of resistance. Above
scenario is quite successful to help understand most of the
properties in SmB$_6$. However, the appearance of resistance
saturation at low temperature challenged such scenario\cite{Cooley}.
To reconcile this discrepancy, ``in-gap'' state is invoked to
account for the saturated resistance at low temperature, whose
existence has been revealed by many other techniques\cite{Nanba, Nyhus,
Alekseev, Takigawa, Miyazaki}. But the exact nature of the
``in-gap'' state is hitherto conundrum.

\begin{figure}[t]
\includegraphics[width=0.45\textwidth]{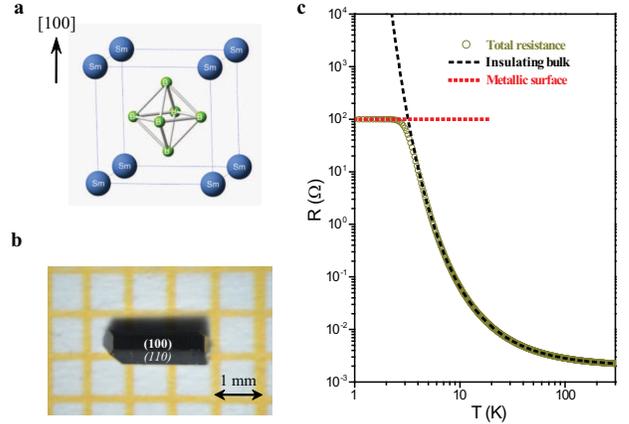}
\caption{(color online). (a) The CsCl-type structure of SmB$_6$ with P$_{m3m}$ space group. Sm ions and B$_6$ octahedron are located at the corner and center of the cubic lattice, respectively. (b) The picture of SmB$_6$ single crystal with both (100) and (110) surfaces. (c) Temperature-dependent resistance for idea topological insulator. The black dashed line represents insulating bulk contribution to total resistance due to thermal activation. The red dashed line represents metallic surface state contribution to total resistance. The green hollow circles is the total resistance containing contributions from both insulating bulk and metallic surface state.}
\end{figure}

Recently, theoretical progress on topological Kondo insulator (TKI) paved a new way to decipher the
resistance saturation in SmB$_6$\cite{Dzero, Dzero1, Alexandrov,
Takimoto, Lu} which was attributed to nontrivial topological surface
state (TSS) as shown in Fig. 1(c). In several recent experiments performed on SmB$_6$, the pronounced surface-dominated transport has been observed \cite{Kim,
Kim1, Kim2, Wolgast, Zhang}. Moreover, angle-resolved photoemission spectroscopy (ARPES) and
quantum oscillation experiments showed direct evidence for
a two-dimensional (2D) Fermi surface\cite{Miyazaki, Jiang, Neupane,
Xu, Li} for SmB$_6$. Very recent spin-resolved ARPES experiment further supports the nontrivial topological nature of above surface state\cite{Xu2}. In this letter, we report magnetoresistance studies on SmB$_6$ single crystals up to 55 Tesla. Our results on temperature- and angular-dependent magnetoresistance offer a new piece of evidence on surface state below 5~K. Moreover, a field-dependent bulk gap with critical magnetic field (H$_c$) of 196 T is extracted from our temperature-dependnet magnetoresistance under different magnetic fields. Finally, a consistent picture to understand the low-temperature transport with both bulk and surface state in SmB$_6$ is proposed.

\begin{figure}[t]
\includegraphics[width=0.45\textwidth]{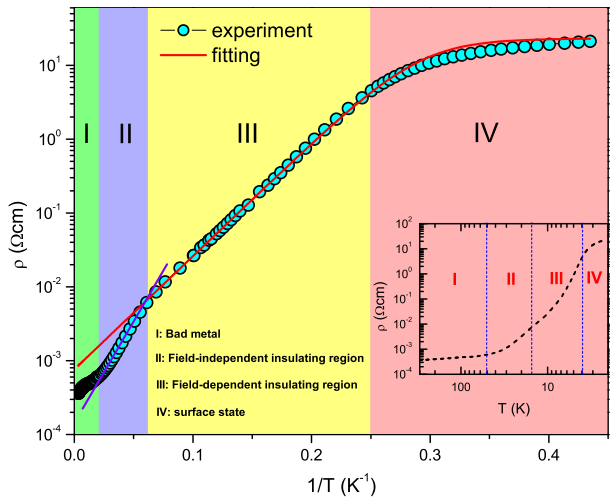}
\caption{(color online). Temperature-dependent resistivity of SmB$_6$ shown in Arrhenius plot. The red solid line is the fitting result for region III and IV with formula 1/$\rho$ = 1/$\rho_0$$e^{-\frac{\Delta_{III}}{k_BT}}$ + 1/$\rho_s$, where $\rho_0$$e^{-\frac{\Delta_{III}}{k_BT}}$ is dominated insulating contribution in region III and $\rho_s$ is the dominated surface state contribution in region IV. The pink solid line is a guiding line to show Arrhenius law for region II. The inset is the temperature-dependent resistivity of SmB$_6$ shown in logarithmic plot.}
\end{figure}

\begin{figure}[t]
\centering
\includegraphics[width=0.45\textwidth]{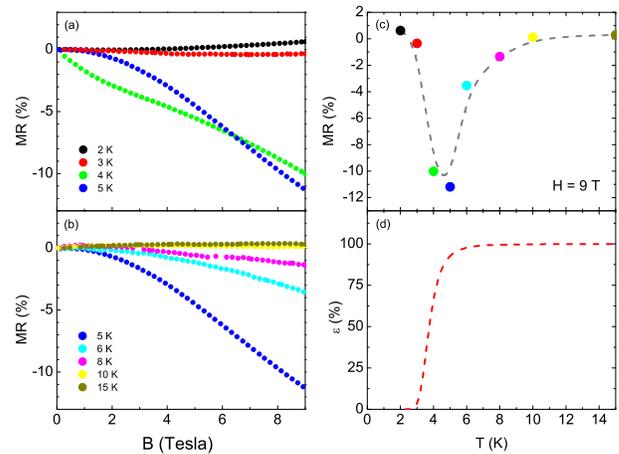}
\caption{(color online). Magnetic-field dependent magnetoresistance of SmB$_6$ with H $\perp$ (100) surface taken (a) below 5~K and (b) between 5 and 15~K. (c) Temperature-dependent magnetoresistance of SmB$_6$ under H $\perp$ (100) = 9~T. The colored circles stand for the value of magnetoresistance taken under H = 9~T at different temperatures, and the black dashed line is the guide for eyes. (d) Temperature-dependent proportion of the bulk state contribution to magnetoresistance from two-channels model fitting in Fig.2. The surface state is supposed to have no field effect on resistance and the negative magnetoresistance is only from bulk state contribution. In this case, we have $\varepsilon$ = ($\rho$/$\rho_0$$e^{-\frac{\Delta}{k_BT}}$)$^2$.}
\end{figure}

For an ideal topological insulator (TI), the temperature-dependent resistance at high temperature is dominated by bulk insulating gap. With decreasing temperature, it will be eventually short-circuited by contribution from surface state as shown in Fig.~1(c). For SmB$_6$, as shown in the inset of Fig.~2, its temperature-dependent resistivity roughly resembles the behavior for an idea TI. However, unlike the ideal TI, the insulating gap in SmB$_6$ opens only below T$_K$ and shows a two-gap behavior. Above T$_K$, considering the incoherent scattering of 5d conduction electrons from local 4f electron, a bad metal behavior is expected for this system. Below T$_K$, there are two different insulating regions with energy gap $\Delta_{II}$ $\sim$ 65 K for region II and $\Delta_{III}$ $\sim$ 35 K for region III observed in temperature-dependent resistivity curve as shown in Fig.~2, which is consistent with previous result\cite{Sluchanko}. Although such two-gap behavior has been widely studied, the origin is still an open issue\cite{Sluchanko}. Here, we observed a different magnetic field effect on above two insulating regions which would be helpful to understand its origin. Below 16~K, the magnetoresistance (MR) in Fig.~3(b) shows a crossover behavior from a faint positive MR to a pronounced negative MR and its amplitude is gradually enhanced with decreasing temperature. Usually, a positive MR is expected for conventional semiconductor but if we consider field effect on Kondo gap in our case, a negative MR is also not surprised. However, it is surprising that region II and region III has different sign for MR, suggesting distinct nature of insulating behavior or gap in these two regions. In present study, we would only focus on the region III with field-dependent insulating behavior. As further temperature decreasing below 5~K, the amplitude of negative MR promptly decreases and then its sign also changes at 2~K with a small MR as shown in Fig.~3(a). It's interesting to plot the amplitude of MR taken under H = 9~T at various temperatures as shown in Fig.~3(c). A crossover behavior is observed with an inflexion around 5~K, consistent with resistance saturation shown in Fig.~2, indicating the same origin for both crossover behaviors. As we know, the resistance saturation is recently ascribed to metallic surface state. We would like to follow this line to explain our MR behavior. As shown in Fig.2, we firstly use two-channels model for idea TI to fit our temperature-dependent resistance in region III and IV, in which both of insulating bulk state and metallic surface state are involved. Then, considering the small MR at 2K, we make an zero-MR approximation for metallic surface state. Finally, we could derive the proportion of bulk contribution to total MR as shown in Fig.3(d). It's very clear that the temperature-dependent MR in Fig.3(c) is perfectly consistent with the crossover from bulk to surface dominated transport.

\begin{figure*}[t]
\centering
\includegraphics[width=0.8\textwidth]{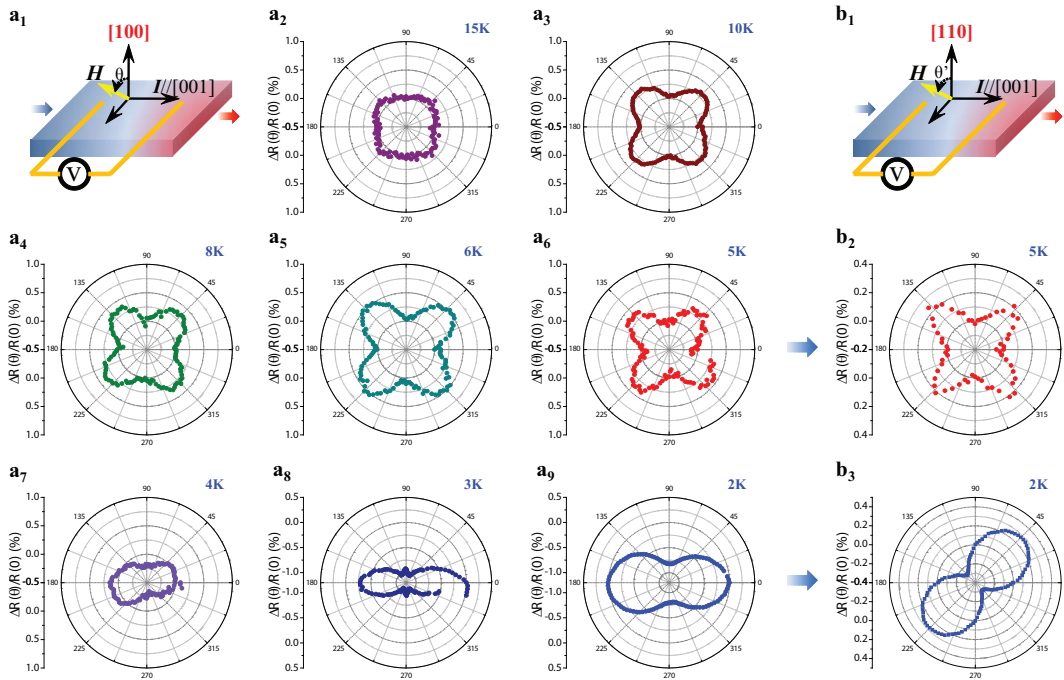}
\caption{(color online). (a$_1$) The electric current with 100
$\mu$A is applied along the [001] direction of sample with (100) surface
and the magnetic field is rotating in the plane perpendicular to the
current direction. (a$_2$)-(a$_9$) Temperature-dependent AMR
patterns of SmB$_6$ sample with (100) surface under 9 Tesla external
magnetic field (H). (b$_1$) The electric current with 100 $\mu$A is
applied along the [001] direction of sample with (110) surface and the
magnetic field is rotating in the plane perpendicular to the current
direction. (b$_2$, b$_3$) AMR patterns at 2 and 5~K for SmB$_6$
single crystal with (110) surface under H = 9~Tesla. R(0) is about 200.3 and 339.7 $\Omega$ at 2K for the (100) and (110) surfaces, respectively. Note that $\theta$' = $\theta$-$\frac{\pi}{4}$. The angular coordinate in all graphs is $\theta$.
}
\end{figure*}

In order to further confirm the crossover from bulk to surface dominated transport, we measured the angular-dependent magnetoresistance (AMR). Since AMR is a powerful tool to reveal the dimensionality of charge transport, the temperature-dependent AMR measurement is performed on SmB$_6$ single crystals with different crystal faces. In Fig.~4(a), a very remarkable evolution of AMR pattern from four-fold to two-fold symmetry was observed below 5~K, consistent with above MR results. Considering the bulk cubic symmetry as shown in Fig.~1(a), the four-fold symmetry in AMR pattern above 5 K was compatible with bulk symmetry and could be attributed to bulk contribution. But the two-fold symmetry in AMR pattern below 5~K was completely incompatible with bulk symmetry. Two possibilities could contribute to such two-fold symmetry. One is the bulk symmetry breaking and the other is the emergence of 2D surface state. In order to distinguish above two possibilities, we further measured the AMR pattern on samples with (100) and (110) surfaces respectively and the current was applied along the same [001] direction. If the two-fold symmetry arises from bulk symmetry breaking, the AMR pattern with the same current orientation should be independent on the choice of surface to make electrodes. Otherwise, it originates from 2D surface state. As shown in Fig.~4(b), a $\frac{\pi}{4}$ phase difference was observed between AMR patterns with different surface crystal faces at 2~K and it disappeared at 5~K with the appearance of four-fold symmetry. The $\frac{\pi}{4}$ is exactly the angle between (100) and (110) planes. It indicates that the observed two-fold symmetry in AMR patterns are dependent on surface crystal face. Therefore, the possibility of bulk symmetry breaking could be excluded and a 2D surface state is validated by our AMR results. This result is consistent with previous transport experiment with different sample size\cite{Kim1, Kim2}, in which a size-independent surface contribution is observed. Thus, the two-fold AMR pattern observed in present study shows the same ability as sample size effect to prove surface state in SmB$_6$. Moreover, the evolution of temperature-dependent AMR further confirms the crossover from bulk to surface dominated transport around 5~K.

In TKI scenario for SmB$_6$, the formation of Kondo gap should be the origin for metallic surface state. Thus, study on the Kondo gap and its correlation with surface state would be also very important to elucidate the origin of surface state observed in various techniques. As we discussed before, Kondo gap is sensitive to magnetic field and could be collapsed under extremely high magnetic field. This offers an opportunity to study the Kondo gap evolution with changing magnetic field. As shown in Fig.5, we expand our measurement on temperature-dependent resistance to high magnetic field up to 55~Tesla. Under different magnetic fields, the temperature-dependent resistances show a similar behavior consistent with two-channels model described before. The remarkable negative MR only appears in intermediate temperature region. Although the MR at 2K under high magnetic field becomes negative again, the amplitude of MR is still much smaller than that in intermediate temperature region, consistent with a zero-MR approximation. Using two-channels model, we could extract the field-dependent insulating gap in region III from our data as shown in the inset of Fig.5. The insulating gap is continuously suppressed with increasing magnetic field. If we use linear extrapolation to get the critical field to collapse the insulating gap, the insulating gap would be closed under the critical magnetic field of about 196~Tesla. Such gap collapse is expected for Kondo insulator and has been observed in several Kondo insulators, such as YbB$_{12}$\cite{Sugiyama} and Ce$_3$Bi$_4$Pt$_3$\cite{Jaime}. In YbB$_{12}$, similar gap collapse in transport has been already studied and the critical field from MR is about 45~Tesla which is in agreement with the expected value from Kondo gap ($\Delta_K$) using H$_c$ = $\frac{\Delta_K}{gJ\mu_B}$\cite{Sugiyama}. Here, we could also estimated the critical field under Kondo gap scenario. Since the theoretical $gJ$ value ($gJ$ = 0.476) for SmB$_6$ is much smaller than that for YbB$_{12}$ ($gJ$ = 1.9), a much higher critical field would be expected under the same Kondo gap. The estimated critical field for SmB$_6$ is about 178~Tesla in agreement with experimental value of 196~Tesla within experimental uncertainty. Here, the Kondo gap $\Delta_K$ is defined as 2$\Delta_{III}$ = 57K. Our result strongly supports that the insulating behavior in region III is produced by the formation of a field-dependent Kondo gap. Surface state contribution in transport following such insulating behavior should be related to Kondo gap, supporting TKI scenario. Besides, no SdH oscillation has been observed in present high-field study, which might be due to the relatively low mobility as reported before\cite{Syers}. Finally, we would emphasize that the insulating gap in region II looks different from Kondo gap and it shows a very weak field-dependence in present study. Although a clear spin gap behavior was observed in region II by previous NMR experiment, there is also absence of any field effect for such spin gap\cite{Takigawa,Caldwell}. How to understand it needs further investigation.

\begin{figure}[t]
\centering
\includegraphics[width=0.45\textwidth]{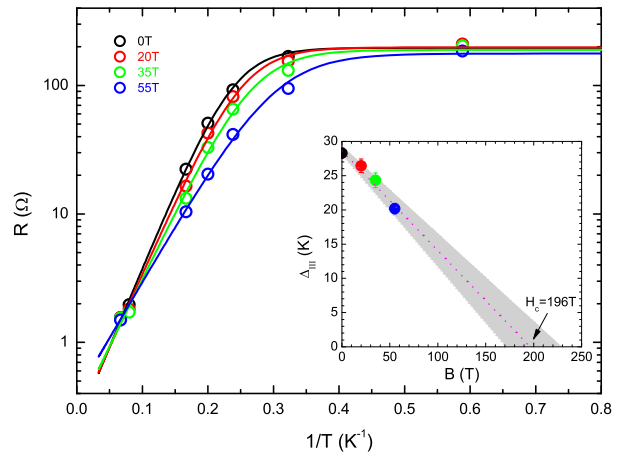}
\caption{(color online). Temperature-dependent resistances of SmB$_6$ shown in arrhenius plot under different magnetic fields. The solid lines are the two-channels model fitting results with formula 1/$\rho$ = 1/$\rho_0$$e^{-\frac{\Delta_{III}}{k_BT}}$ + 1/$\rho_s$, where $\rho_0$$e^{-\frac{\Delta_{III}}{k_BT}}$ is dominated insulating contribution in region III and $\rho_s$ is the dominated surface state contribution in region IV. The inset is the field-dependent energy gap derived from above two-channels model fittings. The dotted line is the linear extrapolation which gives a critical field of about 196 T to collapse the energy gap. The gray area represent the maximum uncertainty of above linear extrapolation}
\end{figure}

In conclusion, we found that both temperature-depedent and angular-dependent MR in SmB$_6$ single crystal exhibit a similar crossover behavior around 5~K. By performing surface-dependent AMR, we gave unambiguous evidence on 2D surface state and confirm such metallic surface state as the origin of above crossover behavior, which is consistent with previous results on surface state. Furthermore, we also reveal that the insulating behavior in SmB$_6$ below 16~K is sensitive to magnetic field which is in agreement with Kondo gap scenario. Finally, Our results give a consistent picture to understand the low-temperature transport behavior in SmB$_6$, which strongly supports TKI scenario.

The authors are grateful for the discussions with Z. Sun, D. L.
Feng, L. Li, S. Y. Li, Y. Y. Wang, G. M. Zhang, X. Dai and Z. Fang.
This work is supported by the National Natural Science Foundation of
China (Grants No. 11190021), the "Strategic Priority Research Program (B)" of the Chinese Academy of
Sciences (Grant No. XDB04040100), the National Basic Research
Program of China (973 Program, Grant No. 2012CB922002), Research Fund for the Doctoral Program of Higher
Education of China(Grant No.20133402110032), China Postdoctoral
Science Foundation funded project(Grant No. 2013M531508) and the
Chinese Academy of Sciences. T. Wu acknowledges Recruitment Program of Global Experts and CAS Hundred Talent Program.

\end{document}